\documentclass[11pt,a4paper]{article}

\usepackage{amssymb}

\usepackage[dvips]{graphicx}

\begin{document}

\title{\bf Scheme independent consequence of the NSVZ relation for ${\cal N}=1$
SQED with $N_f$ flavors}

\author{
A.L.Kataev\\
{\small{\em Institute for Nuclear Research of the Russian Academy of Science,}}\\
{\small {\em 117312, Moscow, Russia}},\\
\\
K.V.Stepanyantz\\
{\small{\em Moscow State University}}, {\small{\em  Physical
Faculty, Department  of Theoretical Physics}}\\
{\small{\em 119991, Moscow, Russia}}}

\maketitle

\begin{abstract}
The exact NSVZ $\beta$-function is obtained for ${\cal N}=1$ SQED
with $N_f$ flavors in all orders of the perturbation theory, if
the renormalization group functions are defined in terms of the
bare coupling constant and the theory is regularized by higher
derivatives. However, if the renormalization group functions are
defined in terms of the renormalized coupling constant, the NSVZ
relation between the $\beta$-function and the anomalous dimension
of the matter superfields is valid only in a certain (NSVZ)
scheme. We prove that for ${\cal N}=1$ SQED with $N_f$ flavors the
NSVZ relation is valid for the terms proportional to $(N_f)^1$ in
an arbitrary subtraction scheme, while the terms proportional to
$(N_f)^k$ with $k\ge 2$ are scheme dependent. These results are
verified by an explicit calculation of a three-loop
$\beta$-function and a two-loop anomalous dimension made with the
higher derivative regularization in the NSVZ and MOM subtraction
schemes. In this approximation it is verified that in the MOM
subtraction scheme the renormalization group functions obtained
with the higher derivative regularization and with the dimensional
reduction coincide.
\end{abstract}

\unitlength=1cm

Keywords: higher covariant derivative regularization,
supersymmetry, $\beta$-function, subtraction scheme.


\section{Introduction}
\hspace{\parindent}

The $\beta$-function of ${\cal N}=1$ supersymmetric gauge theories
is related with the anomalous dimension of the matter superfields.
This relation, derived in Refs.
\cite{Novikov:1983uc,Novikov:1985rd,Shifman:1986zi}, is usually
called "the exact Novikov, Shifman, Vainshtein, and Zakharov
(NSVZ) $\beta$-function". For the ${\cal N}=1$ supersymmetric
Yang--Mills theory without matter superfields the NSVZ
$\beta$-function was obtained in Refs.
\cite{Novikov:1983uc,Jones}. In the case of the ${\cal N}=1$
supersymmetric electrodynamics (SQED) with $N_f$ flavors, which is
considered in this paper, the NSVZ $\beta$-function has the
following form \cite{Vainshtein:1986ja,Shifman:1985fi}:

\begin{equation}\label{NSVZ}
\beta(\alpha_0) = \frac{\alpha_0^2
N_f}{\pi}\Big(1-\gamma(\alpha_0)\Big).
\end{equation}

\noindent This relation, derived from general arguments, can be
verified by explicit calculations. Usually for calculating quantum
corrections SUSY theories are regularized by the dimensional
reduction (DRED) \cite{Siegel:1979wq} supplemented by the
$\overline{\mbox{DR}}$-scheme. However, DRED is not mathematically
consistent \cite{Siegel:1980qs}. As a consequence, supersymmetry
can be broken by quantum corrections in higher loops
\cite{Avdeev:1982np,Avdeev:1982xy,Velizhanin:2008rw}.

Explicit calculations made in the $\overline{\mbox{DR}}$-scheme in
the one- \cite{Ferrara:1974pu} and two-loop \cite{Jones:1974pg}
approximations agree with the NSVZ $\beta$-function, because a
two-loop $\beta$-function and a one-loop anomalous dimension are
scheme independent in theories with a single coupling constant. In
higher orders
\cite{Avdeev:1981ew,Jack:1996vg,Jack:1996cn,Jack:1998uj,Harlander:2006xq,Jack:2007ni}
the exact NSVZ relation for the renormalization group (RG)
functions defined in terms of the renormalized coupling constant
can be obtained with the $\overline{\mbox{DR}}$-scheme after an
additional finite renormalization. This finite renormalization
should be fixed in each order of the perturbation theory, starting
from the three-loop approximation. However, at present, there are
no general prescriptions, how one should construct this finite
renormalization using the $\overline{\mbox{DR}}$-scheme in all
orders.

In the Abelian case the NSVZ $\beta$-function can be obtained in
all orders using the Slavnov higher derivative (HD) regularization
\cite{Slavnov:1971aw,Slavnov:1972sq}. This regularization is
mathematically consistent and can be formulated in an explicitly
supersymmetric way \cite{Krivoshchekov:1978xg,West:1985jx}. The HD
regularization allows to obtain the NSVZ $\beta$-function for the
RG functions defined in terms of the bare coupling constant
\cite{Stepanyantz:2011jy,Kataev:2013eta}. The reason for this is
that the integrals needed for obtaining such a $\beta$-function in
SUSY theories are integrals of total derivatives
\cite{Soloshenko:2003nc,Pimenov:2009hv,Stepanyantz:2011zz} and
even double total derivatives
\cite{Smilga:2004zr,Stepanyantz:2011bz,Stepanyantz:2012zz,Stepanyantz:2012us}.
As a consequence, one of the loop integrals can be calculated
analytically, and a $\beta$-function in an $L$-loop approximation
can be related with an anomalous dimension of the matter
superfields in the $(L-1)$-loop approximation
\cite{Stepanyantz:2011jy}. However, if the RG functions are
defined in terms of the renormalized coupling constant, the NSVZ
$\beta$-function is obtained only in a special subtraction scheme.
This scheme was constructed in \cite{Kataev:2013eta} by imposing
the boundary conditions

\begin{equation}\label{NSVZ_Scheme}
Z_3(\alpha,x_0) = 1; \qquad Z(\alpha,x_0)=1,
\end{equation}

\noindent where $x_0$ is a certain value of $x=\ln\Lambda/\mu$.
Without loss of generality it is possible to choose $x_0=0$.

In this paper we study ${\cal N}=1$ SQED with $N_f$ flavors. It is
shown that the coefficients of the anomalous dimension of the
matter superfields proportional to $(N_f)^0$ and the coefficients
of the $\beta$-function proportional to $(N_f)^1$ are scheme
independent in all orders. As a consequence, they satisfy the NSVZ
relation in all orders independently of a choice of a subtraction
scheme. In order to verify this result we explicitly calculate a
three-loop $\beta$-function and a two-loop anomalous dimension
using the HD regularization with different renormalization
prescriptions. Also we present the results of a similar
calculation \cite{Jack:1996vg} which is made using the
$\overline{\mbox{DR}}$-scheme. Then it is explicitly demonstrated
that the terms proportional to $(N_f)^1$ in the $\beta$-function
and to $(N_f)^0$ in the anomalous dimension are scheme independent
and satisfy the NSVZ relation.

The paper is organized as follows: In Sect.
\ref{Section_Regularization} we remind how the NSVZ
$\beta$-function can be obtained for ${\cal N}=1$ SQED with $N_f$
flavors using the HD regularization for the RG functions defined
in terms of the bare coupling constant. The standard definition of
the RG functions (in terms of the renormalized coupling constant)
and their scheme dependence are discussed in Sect.
\ref{Section_Scheme_Dependence}. In Sect.
\ref{Section_Explicit_Three_Loop} the results are verified by an
explicit three-loop calculation. The two-loop anomalous dimension
of the matter superfields and the three-loop $\beta$-function in
different subtraction schemes are compared in Sect.
\ref{Section_DRED_NSVZ}.

\section{The NSVZ $\beta$-function for ${\cal N}=1$ SQED
with $N_f$ flavors}
\hspace{\parindent}\label{Section_Regularization}

In terms of ${\cal N}=1$ superfields ${\cal N}=1$ SQED with $N_f$
flavors in the massless limit is described by the action

\begin{equation}\label{Action}
S = \frac{1}{4e_0^2}\mbox{Re}\int d^4x\,d^2\theta\,W^a W_a +
\sum\limits_{i=1}^{N_f} \frac{1}{4} \int
d^4x\,d^4\theta\,\Big(\phi_i^* e^{2V}\phi_i + \widetilde\phi_i^*
e^{-2V} \widetilde\phi_i\Big),
\end{equation}

\noindent where $e_0$ is a bare coupling constant. The exact NSVZ
$\beta$-function can be naturally obtained for this theory, if the
HD method is used for a regularization.

In order to regularize this theory by higher derivatives, it is
necessary to insert into the first term of Eq. (\ref{Action}) a
regularizing function $R(\partial^2/\Lambda^2)$ such that $R(0)=1$
and $R(\infty)=\infty$ \cite{Slavnov:1971aw,Slavnov:1972sq}:

\begin{equation}\label{Regularized_Action}
\frac{1}{4e_0^2}\mbox{Re}\int d^4x\,d^2\theta\,W^a W_a \to
\frac{1}{4e_0^2}\mbox{Re}\int d^4x\,d^2\theta\,W^a
R(\partial^2/\Lambda^2) W_a.
\end{equation}

\noindent It is convenient to choose $R =
1+\partial^{2n}/\Lambda^{2n}$, where $\Lambda$ is a dimensionful
parameter. Also we should insert into the generating functional
the Pauli--Villars determinants, which cancel the remaining
one-loop divergences \cite{Faddeev:1980be,Slavnov:1977zf}. Then
the generating functional can be written in the following form:

\begin{equation}\label{Generating_Functional}
Z[J,j,\widetilde j] = \int DV\, D\phi\,
D\widetilde\phi\,\prod\limits_{I=1}^n (\det(V,M_I))^{c_I
N_f}\exp\Big(i S_{\mbox{\scriptsize reg}} + i S_{\mbox{\scriptsize
gf}} + i S_{\mbox{\scriptsize source}} \Big),
\end{equation}

\noindent where $M_I = a_I\Lambda$ are masses of the
Pauli--Villars superfields and the coefficients $a_I$ do not
depend on the bare charge. $S_{\mbox{\scriptsize reg}}$ is the
regularized action containing the HD term and
$S_{\mbox{\scriptsize gf}}$ is the gauge fixing term. In the
Abelian case it is not necessary to introduce ghost (super)fields.
The Pauli--Villars determinants $\det(V,M_I)$ are constructed
exactly as in the case $N_f=1$ (see, e.g. Refs.
\cite{Stepanyantz:2011jy,Kataev:2013eta}). For cancelation of
remaining one-loop divergences the coefficients $c_I$ should
satisfy the conditions $ \sum_I c_I = 1$ and $\sum_I c_I M_I^2 =0$
\cite{Slavnov:1977zf}.

Let us consider a part of the effective action corresponding to
the two-point functions of the gauge and matter superfields:

\begin{eqnarray}\label{DG_Definition}
&& \Gamma^{(2)} - S_{\mbox{\scriptsize gf}} = - \frac{1}{16\pi}
\int \frac{d^4p}{(2\pi)^4}\, d^4\theta\, V(\theta,-p)\,
\partial^2\Pi_{1/2} V(\theta,p)\,
d^{-1}(\alpha_0,\Lambda/p)\nonumber\\
&& + \frac{1}{4} \sum\limits_{i=1}^{N_f} \int
\frac{d^4p}{(2\pi)^4} d^4\theta\, \Big(\phi_i^*(\theta,-p)\,
\phi_i(\theta,p) + \widetilde\phi_i^*(\theta,-p)\,
\widetilde\phi_i(\theta,p) \Big) G(\alpha_0,\Lambda/p),
\end{eqnarray}

\noindent where $\partial^2\Pi_{1/2}  = - D^a \bar D^2 D_a/8$ is
the supersymmetric transversal projector. The NSVZ relation is
naturally obtained for the RG functions defined according to the
following prescriptions:

\begin{eqnarray}\label{Beta_Definition1}
&& \beta\Big(\alpha_0(\alpha,\Lambda/\mu)\Big) \equiv \frac{d
\alpha_0(\alpha,\Lambda/\mu)}{d\ln\Lambda}
\Big|_{\alpha=\mbox{\scriptsize const}};\vphantom{\Bigg|}\\
\label{Gamma_Definition1} &&
\gamma\Big(\alpha_0(\alpha,\Lambda/\mu)\Big) \equiv - \frac{d \ln
Z(\alpha,\Lambda/\mu)}{d
\ln\Lambda}\Big|_{\alpha=\mbox{\scriptsize const}},
\end{eqnarray}

\noindent where $\alpha$ is the renormalized coupling constant and
$Z$ is the renormalization constant for the matter superfields.
They can be found by requiring finiteness of the functions
$d^{-1}(\alpha_0(\alpha,\Lambda/\mu),\Lambda/p)$ and $Z
G(\alpha,\Lambda/\mu,\Lambda/p)$ in the limit $\Lambda\to \infty$.
Certainly, the renormalized coupling constant $\alpha$ and the
renormalization constant $Z$ are not uniquely defined and depend
on a choice of a renormalization scheme \cite{Vladimirov:1975mx}.
However, it is possible to prove (see e.g. \cite{Kataev:2013eta})
that the RG functions (\ref{Beta_Definition1}) and
(\ref{Gamma_Definition1}) are independent of a renormalization
prescription.

If the HD method is used for a regularization, the integrals which
determine the $\beta$-function (\ref{Beta_Definition1}) are
integrals of (double) total derivatives
\cite{Soloshenko:2003nc,Smilga:2004zr,Stepanyantz:2011jy}.
Therefore, one of the loop integrals can be calculated
analytically giving the relation \cite{Stepanyantz:2011jy}

\begin{equation}\label{Green_Function_Relation}
\frac{\beta(\alpha_0)}{\alpha_0^2} = \frac{d}{d\ln \Lambda}\,
\Big(d^{-1}(\alpha_0,\Lambda/p)-\alpha_0^{-1}\Big)\Big|_{p=0} =
\frac{N_f}{\pi}\Big(1-\frac{d}{d\ln\Lambda} \ln
G(\alpha_0,\Lambda/q)\Big|_{q=0}\Big) = \frac{N_f}{\pi}\Big(1 -
\gamma(\alpha_0)\Big),
\end{equation}

\noindent which is exact in all orders. Thus, the exact NSVZ
$\beta$-function (\ref{NSVZ}) is obtained for the RG functions
defined in terms of the bare charge independently of a
renormalization prescription.

\section{Scheme dependence of the RG functions defined
in terms of the renormalized coupling constant}
\hspace{\parindent}\label{Section_Scheme_Dependence}

Although the exact NSVZ relation for the considered theory is
naturally obtained for the RG functions defined in terms of the
bare coupling constant, usually the RG functions are defined in a
different way:

\begin{eqnarray}
\label{Beta_Definition2} &&
\widetilde\beta\Big(\alpha(\alpha_0,\Lambda/\mu)\Big) \equiv
\frac{d\alpha(\alpha_0,\Lambda/\mu)}{d\ln\mu}\Big|_{\alpha_0=\mbox{\scriptsize const}};\\
\label{Gamma_Definition2} &&
\widetilde\gamma\Big(\alpha(\alpha_0,\Lambda/\mu)\Big) \equiv
\frac{d}{d\ln\mu}\ln
ZG(\alpha_0,\Lambda/\mu)\Big|_{\alpha_0=\mbox{\scriptsize const}}
= \frac{d\ln Z(\alpha(\alpha_0,\Lambda/\mu),
\Lambda/\mu)}{d\ln\mu}\Big|_{\alpha_0=\mbox{\scriptsize
const}},\qquad
\end{eqnarray}

\noindent where $\alpha_0$ and $\mu$ are considered as independent
variables. By definition, these RG functions depend on the
renormalized coupling constant. Unlike the RG function
(\ref{Beta_Definition1}) and (\ref{Gamma_Definition1}), they
depend on an arbitrariness of choosing $\alpha$ and $Z$.
Therefore, in general, these functions do not satisfy the NSVZ
relation, which was originally derived for the bare quantities.
Nevertheless, as was shown in Ref. \cite{Kataev:2013eta}, if there
is a point $x_0 =\ln\Lambda/\mu_0$ for which the boundary
conditions (\ref{NSVZ_Scheme}) are valid, the RG functions
(\ref{Beta_Definition2}) and (\ref{Gamma_Definition2}) coincide
with the RG functions (\ref{Beta_Definition1}) and
(\ref{Gamma_Definition1}), respectively, and, as a consequence,
satisfy the NSVZ relation.

Under a finite renormalization

\begin{equation}
\alpha \to \alpha'(\alpha);\qquad Z'(\alpha',\Lambda/\mu) =
z(\alpha) Z(\alpha,\Lambda/\mu)
\end{equation}

\noindent the $\beta$-function (\ref{Beta_Definition2}) and the
anomalous dimension (\ref{Gamma_Definition2}) are changed as
follows:

\begin{eqnarray}\label{Beta_Transformation}
&& \widetilde \beta'(\alpha') =
\frac{d\alpha'}{d\ln\mu}\Big|_{\alpha_0=\mbox{\scriptsize const}}
= \frac{d\alpha'}{d\alpha} \widetilde
\beta(\alpha);\\
\label{Gamma_Transformation} && \widetilde \gamma'(\alpha') =
\frac{d\ln Z'}{d\ln\mu}\Big|_{\alpha_0=\mbox{\scriptsize const}} =
\frac{d\ln z}{d\alpha}\cdot \widetilde \beta(\alpha) +
\widetilde\gamma(\alpha).
\end{eqnarray}

\noindent Using these equations it is easy to see that if
$\widetilde\beta(\alpha)$ and $\widetilde\gamma(\alpha)$ satisfy
the NSVZ relation, then

\begin{equation}\label{New_NSVZ}
\widetilde\beta'(\alpha') = \frac{d\alpha'}{d\alpha}\cdot
\frac{\alpha^2 N_f}{\pi}\cdot
\frac{1-\widetilde\gamma'(\alpha')}{1- \alpha^2 N_f (d\ln z/d
\alpha)/\pi}\,\Big|_{\alpha=\alpha(\alpha')}.
\end{equation}

\noindent This result generalizes a similar equation presented in
Ref. \cite{Kataev:2013eta} for the case $N_f=1$.

Let us note that quantum corrections to the coupling constant are
produced by diagrams which contain at least one loop of the matter
superfields. Such a loop gives a factor $N_f$. Thus, it is
reasonable to make finite renormalizations of the coupling
constant proportional to $N_f$:

\begin{equation}\label{Finite_Renormalization}
\alpha'(\alpha) - \alpha = O(N_f);\qquad z(\alpha) =
O\left((N_f)^0\right).
\end{equation}

\noindent Then from Eq. (\ref{New_NSVZ}) we see that all scheme
dependent terms in the $\beta$-function are proportional at least
to $(N_f)^2$ in all orders of the perturbation theory. Similarly,
from Eq. (\ref{Gamma_Transformation}) it is evident that the terms
proportional to $(N_f)^0$ in the anomalous dimension are scheme
independent. Also we know that the NSVZ scheme exists. Therefore,
the NSVZ relation is satisfied for terms proportional to $(N_f)^1$
in all orders, while terms proportional to $(N_f)^\alpha$ with
$\alpha\ge 2$ are scheme dependent.

\section{Scheme dependence in the three-loop approximation}
\hspace{\parindent}\label{Section_Explicit_Three_Loop}

In the case of using the HD regularization with
$R=1+\partial^{2n}/\Lambda^{2n}$ for the considered theory the
functions $d^{-1}$ and $G$ in the three- and two-loop
approximations, respectively, are given by the following
expressions:

\begin{eqnarray}
&&\hspace*{-6mm} d^{-1}(\alpha_0,\Lambda/p) = \frac{1}{\alpha_0} +
\frac{N_f}{\pi}\Big(\ln\frac{\Lambda}{p} + d_1\Big) +
\frac{\alpha_0 N_f}{\pi^2}\Big(\ln\frac{\Lambda}{p} + d_2\Big) +
\frac{\alpha_0^2 N_f}{\pi^3}\Big(-\frac{N_f}{2}\ln^2
\frac{\Lambda}{p} + d_3 - \ln\frac{\Lambda}{p}\nonumber\\
&&\hspace*{-6mm} \times\Big(N_f \sum\limits_{I=1}^n c_I \ln a_I +
N_f + \frac{1}{2} + N_f d_2\Big)\Big) + \mbox{(terms vanishing in
the limit $\Lambda\to \infty$)} +
O(\alpha_0^3);\vphantom{\Big(}\\
&& \nonumber\\
&&\hspace*{-6mm} G(\alpha_0,\Lambda/p) = 1 - \frac{\alpha_0}{\pi}
\ln\frac{\Lambda}{p} -\frac{\alpha_0}{2\pi} +
\frac{\alpha_0^2(N_f+1)}{2\pi^2} \ln{}^2 \frac{\Lambda}{p} +
\frac{\alpha_0^2}{\pi^2}\ln\frac{\Lambda}{p}\Big(N_f
\sum\limits_{I=1}^n c_I \ln a_I +\frac{3N_f}{2} +1\Big)\qquad \nonumber\\
&&\hspace*{-6mm} + \frac{\alpha_0^2}{\pi^2} c_2 + \Big(\mbox{terms
vanishing in the limit $\Lambda\to \infty$}\Big) + O(\alpha_0^3),
\end{eqnarray}

\noindent where $d_1$, $d_2$, $d_3$, and $c_2$ are finite
constants, which should be found by explicit calculating Feynman
graphs. These equations are derived similar to the case of ${\cal
N}=1$ SQED, which have been described in details in Ref.
\cite{Kataev:2013eta}. The loop integrals which determine these
Green functions for $N_f=1$ can be found in Refs.
\cite{Soloshenko:2003nc} and \cite{Soloshenko:2003sx}. The
coefficients $d_1$ and $d_2$, which are needed in this paper, are
calculated as follows:

According to Ref. \cite{Soloshenko:2003nc} the function
$d^{-1}(\alpha_0,\Lambda/p)$ (obtained with the HD regularization)
in the two-loop approximation is given by

\begin{equation}\label{Invariant_Charge}
d^{-1}(\alpha_0,\Lambda/p) = \frac{1}{\alpha_0} + \frac{N_f}{\pi}
\sum\limits_{I=1}^n c_I \Bigg(\ln \frac{M_I}{p} +
\sqrt{1+\frac{4M_I^2}{p^2}}\, \mbox{arctanh}
\sqrt{\frac{p^2}{4M_I^2+p^2}}\Bigg) + \alpha_0 N_f I_2 +
O(\alpha_0^2),
\end{equation}

\noindent where

\begin{eqnarray}\label{I2}
&& I_2 \equiv 64\pi^2 \int \frac{d^4k}{(2\pi)^4}
\frac{d^4q}{(2\pi)^4}\frac{1}{k^2 R_k} \Bigg\{ \frac{(k+p+q)^2 +
q^2 - k^2 - p^2}{q^2 (q+p)^2 (k+q)^2 (k+q+p)^2} -
\sum\limits_{I=1}^n c_I \frac{1}{\left(q^2+M_I^2\right)}
\qquad\quad\nonumber\\
&& \times \frac{1}{\left((q+p)^2 + M_I^2 \right)\left((k+q)^2 +
M_I^2 \right)}\Bigg(\frac{(k+p+q)^2 + q^2 - k^2 - p^2}{(k+q+p)^2 +
M_I^2} - \frac{4 M_I^2}{q^2+M_I^2}\Bigg) \Bigg\}.
\end{eqnarray}

\noindent This expression is written in the Euclidian space after
the Wick rotation and $R_k \equiv R(k^2/\Lambda^2) = 1+
k^{2n}/\Lambda^{2n}$. Subtracting the term proportional to
$\ln\Lambda/p$ and taking the limit $p\to 0$, from Eq.
(\ref{Invariant_Charge}) we obtain

\begin{equation}\label{D1}
d_1 = \sum\limits_{I=1}^{n} c_I \ln a_I + 1.
\end{equation}

The massive two-loop integrals coming from the Pauli--Villars
determinants are finite in the infrared region. As a consequence,
calculating their sum (which depends only on $p/\Lambda$) it is
possible to set $p=0$. Then the corresponding terms in Eq.
(\ref{I2}) give the vanishing integral of a total derivative

\begin{equation}
64\pi^2 \sum\limits_{I=1}^n c_I \int \frac{d^4k}{(2\pi)^4}
\frac{d^4q}{(2\pi)^4}\frac{1}{k^2 R_k} \frac{\partial}{\partial
q^\mu} \Bigg(\frac{q^\mu}{\left(q^2 + M_I^2\right)^2\left((q+k)^2
+ M_I^2\right)}\Bigg) = 0.
\end{equation}

The remaining integral can be rewritten in the following form:

\begin{eqnarray}\label{Original_Integral}
I_2 = 128\pi^2 \int \frac{d^4k}{(2\pi)^4} \frac{d^4q}{(2\pi)^4}
\frac{q_\mu (q+k+p)_\mu}{k^2 R_k\, q^2 (q+p)^2 (k+q)^2 (k+q+p)^2}
+ o(1),
\end{eqnarray}

\noindent where $o(1)$ denotes terms vanishing in the limit $p\to
0$. Deriving this equation we take into account that the term
proportional to $k_\mu p_\mu$ vanishes, because the sign of this
term is inverted after the change of variables $q_\mu \to q_\mu -
p_\mu$ and the subsequent replacement $p_\mu \to -p_\mu$. In order
to calculate the above integral, we add to it

\begin{equation}
0 = - 128\pi^2 \int \frac{d^4k}{(2\pi)^4} \frac{d^4q}{(2\pi)^4}
\frac{q_\mu (q+k+p)_\mu}{k^2 R_k\, q^4 (k+q+p)^4} +
\frac{1}{\pi^2}\Big(\ln\frac{\Lambda}{p}+ \frac{1}{2}\Big) + o(1).
\end{equation}

\noindent The integral over the loop momentums obtained after this
procedure is convergent in both ultraviolet and infrared regions
and depends only on $p/\Lambda$. Therefore, its value in the limit
$p \to 0$ can be found by setting $\Lambda\to \infty$, so that
$R_k \to 1$. As a consequence,

\begin{equation}
d_2 = \frac{1}{2} - 128 \pi^4 \int \frac{d^4k}{(2\pi)^4}
\frac{d^4q}{(2\pi)^4} \frac{q^\mu (q+k+p)_\mu \left((k^2 + 2
k^\alpha q_\alpha)(p^2 + 2 p^\beta q_\beta) - 2q^2 k^\alpha
p_\alpha \right)}{k^2 q^4 (q+k+p)^4 (q+p)^2 (q+k)^2}.
\end{equation}

\noindent Presenting this integral as a sum of scalar integrals
and calculating them using the dimensional regularization
\cite{'tHooft:1972fi,Bollini:1972ui,Ashmore:1972uj,Cicuta:1972jf}
in the limit $d\to 4$ we obtain

\begin{equation}\label{D2}
d_2 = \frac{3}{2}\Big(1-\zeta(3)\Big).
\end{equation}

\noindent In this expression the term proportional to $\zeta(3)$
comes from a certain 2-loop  scalar master  integral, which has been calculated in
Ref. \cite{Chetyrkin:1980pr} using the Gegenbauer polynomial x-space technique.

The function $d^{-1}$ expressed in terms of the renormalized
coupling constant $\alpha$ is finite in the limit
$\Lambda\to\infty$ if $\alpha$ is related with the bare coupling
constant $\alpha_0=e_0^2/4\pi$ by the equation

\begin{eqnarray}\label{Alpha0}
&& \frac{1}{\alpha_0} = \frac{1}{\alpha} - \frac{N_f}{\pi} \Big(
\ln\frac{\Lambda}{\mu} + b_1\Big) - \frac{\alpha N_f}{\pi^2}
\Big(\ln\frac{\Lambda}{\mu} + b_2\Big) - \frac{\alpha^2
N_f}{\pi^3}\Big(\frac{N_f}{2}\ln^2 \frac{\Lambda}{\mu}
-\ln\frac{\Lambda}{\mu}\Big(N_f \sum\limits_{I=1}^n c_I \ln
a_I\qquad
\nonumber\\
&& + N_f + \frac{1}{2} - N_f b_1\Big) + b_3 \Big) +
O(\alpha^3).\qquad
\end{eqnarray}

\noindent In this equation $b_1$, $b_2$, and $b_3$ are arbitrary
finite constants, which partially define the subtraction scheme.
The coefficients $b_i$ are multiplied by the factor $N_f$
according to Eq. (\ref{Finite_Renormalization}). Similarly,
divergences in the two-point Green function of the matter
superfields can be cancelled by multiplying the function
$G(\alpha_0,\Lambda/p)$ by the renormalization constant $Z$, which
is given by

\begin{eqnarray}\label{Two_Loop_Z}
&& Z = 1 + \frac{\alpha}{\pi}\Big(\ln\frac{\Lambda}{\mu}+g_1\Big)
+\frac{\alpha^2(N_f +1)}{2\pi^2}\ln^2\frac{\Lambda}{\mu}
-\frac{\alpha^2}{\pi^2}\ln\frac{\Lambda}{\mu}\Big(N_f
\sum\limits_{I=1}^n
c_I\ln a_I - N_f b_1 + N_f + \frac{1}{2} \qquad\nonumber\\
&& - g_1 \Big) + \frac{\alpha^2 g_2}{\pi^2} + O(\alpha^3).
\end{eqnarray}

\noindent Here $g_1$ and $g_2$ are again finite constants, which
(together with $b_i$) fix the subtraction scheme in the considered
approximation. It is easy to see that for arbitrary values of
these constants the function $Z G$ is finite in the limit
$\Lambda\to \infty$.

The anomalous dimension (\ref{Gamma_Definition1}) can be found by
differentiating $\ln Z(\alpha,\Lambda/\mu)$ with respect to
$\ln\Lambda$ and writing the result in terms of $\alpha_0$. Then
we obtain

\begin{equation}\label{Gamma_Answer1}
\gamma(\alpha_0) = - \frac{d\ln Z}{d\ln\Lambda} =
-\frac{\alpha_0}{\pi} + \frac{\alpha_0^2}{\pi^2}\Big(N_f
\sum\limits_{I=1}^n c_I \ln a_I + N_f + \frac{1}{2}\Big) +
O(\alpha_0^3).
\end{equation}

\noindent This expression is independent of the finite constants
$g_i$ and $b_i$, which fix  the subtraction scheme.

The anomalous dimension $\widetilde\gamma(\alpha)$ defined by Eq.
(\ref{Gamma_Definition2}) can be constructed similarly. For this
purpose we rewrite $\ln Z$ in terms of $\alpha_0$ using Eq.
(\ref{Alpha0}) and differentiate the result with respect to
$\ln\mu$. Writing the result in terms of $\alpha$ we obtain

\begin{equation}\label{Gamma_Answer2}
\widetilde\gamma(\alpha) = \frac{d\ln Z}{d\ln\mu} =
-\frac{\alpha}{\pi} + \frac{\alpha^2}{\pi^2}\Big(N_f + N_f
\sum\limits_{I=1}^n c_I \ln a_I - N_f b_1 + N_f
g_1+\frac{1}{2}\Big) + O(\alpha^3).
\end{equation}

\noindent Unlike Eq. (\ref{Gamma_Answer1}) this expression depends
on the constants $g_1$ and $b_1$. However, only the terms
proportional to $(N_f)^1$ depend on these parameters, the terms
proportional to $(N_f)^0$ being independent of them.

Differentiating Eq. (\ref{Alpha0}) with respect to $\ln\Lambda$
and writing the result in terms of $\alpha_0$ we obtain the
$\beta$-function defined by Eq. (\ref{Beta_Definition1}):

\begin{equation}\label{Beta_Answer1}
\frac{\beta(\alpha_0)}{\alpha_0^2} = \frac{N_f}{\pi} +
\frac{\alpha_0 N_f}{\pi^2} - \frac{\alpha_0^2 N_f}{\pi^3}\Big(N_f
\sum\limits_{I=1}^n c_I \ln a_I + N_f +\frac{1}{2} \Big) +
O(\alpha_0^3).
\end{equation}

\noindent This $\beta$-function does not depend  on the finite
constants $g_i$ and $b_i$ and is related with the anomalous
dimension (\ref{Gamma_Answer1}) by Eq. (\ref{NSVZ}). The
$\beta$-function (\ref{Beta_Definition2}) is calculated by
re-expressing $\alpha$ in terms of $\alpha_0$ and differentiating
the result with respect to $\ln \mu$:

\begin{equation}\label{Beta_Answer2}
\frac{\widetilde\beta(\alpha)}{\alpha^2} = \frac{N_f}{\pi} +
\frac{\alpha N_f}{\pi^2} - \frac{\alpha^2 N_f}{\pi^3}\Big(N_f
\sum\limits_{I=1}^n c_I \ln a_I + N_f + \frac{1}{2} - N_f b_1 +
N_f b_2\Big) + O(\alpha^3).
\end{equation}

\noindent This equation implies that the terms proportional to
$N_f$ do not depend on the constants $b_i$ and are, therefore,
scheme independent. Moreover, comparing Eqs. (\ref{Gamma_Answer2})
and (\ref{Beta_Answer2}) we see that for the terms proportional to
$(N_f)^1$ the NSVZ relation is satisfied. This result agrees with
the general statement presented above, which follows from Eq.
(\ref{New_NSVZ}) in all orders of the perturbation theory.

\section{Examples: NSVZ, MOM and $\overline{\mbox{DR}}$ schemes}
\label{Section_DRED_NSVZ} \hspace{\parindent}

Let us compare the results of explicit calculations made with
different subtraction schemes, namely, the NSVZ scheme obtained
with the HD regularization \cite{Kataev:2013eta}, the MOM scheme,
and the $\overline{\mbox{DR}}$ scheme. Certainly, any pair of
these schemes can be related by a finite renormalization
\cite{Vladimirov:1975mx}.

With the HD regularization the NSVZ scheme for the RG function
defined in terms renormalized coupling constant is obtained by
imposing the boundary conditions (\ref{NSVZ_Scheme}) on the
renormalization constants. Choosing $x_0=0$, it is easy to see
that in this case

\begin{equation}
g_1=g_2=0;\qquad b_1=b_2=b_3=0
\end{equation}

\noindent and, therefore,

\begin{eqnarray}\label{Gamma_NSVZ}
&& \widetilde\gamma_{\mbox{\scriptsize NSVZ}}(\alpha) =
\gamma(\alpha) = -\frac{\alpha}{\pi} +
\frac{\alpha^2}{\pi^2}\Big(\frac{1}{2} + N_f\sum\limits_{I=1}^n
c_I \ln a_I  + N_f\Big) +
O(\alpha^3);\\
&&\label{Beta_NSVZ} \widetilde\beta_{\mbox{\scriptsize
NSVZ}}(\alpha) = \beta(\alpha) = \frac{\alpha^2 N_f}{\pi}\Big(1+
\frac{\alpha}{\pi} - \frac{\alpha^2}{\pi^2}\Big(\frac{1}{2}+ N_f
\sum\limits_{I=1}^n c_I \ln a_I + N_f \Big) +
O(\alpha^3)\Big).\qquad
\end{eqnarray}

\noindent Thus, in this scheme the NSVZ relation is valid for
terms proportional to both $(N_f)^1$ and $(N_f)^2$. This is in
agreement with the general result that in the scheme defined by
the conditions (\ref{NSVZ_Scheme}) the NSVZ $\beta$-function is
obtained in all orders, if the theory is regularized by HD.

The MOM scheme is defined by the boundary conditions

\begin{equation}\label{MOM}
Z_{\mbox{\scriptsize MOM}}G(\alpha_{\mbox{\scriptsize MOM}},p=\mu)
= 1; \qquad d^{-1}(\alpha_{\mbox{\scriptsize
MOM}},p=\mu)=\alpha_{\mbox{\scriptsize MOM}}^{-1}
\end{equation}

\noindent imposed on the renormalized Green functions. In this
case

\begin{equation}
g_1 = \frac{1}{2};\qquad g_2 = - c_2 + \frac{1}{4} +\frac{N_f}{2}
b_1;\qquad b_1 = d_1;\qquad b_2 = d_2;\qquad b_3 = d_3+ N_f d_1
d_2.
\end{equation}

\noindent Therefore, in the MOM subtraction scheme the constants
$b_i$ and $g_i$ are related with the finite parts of the Green
functions ($c_i$ and $d_i$). Using Eq. (\ref{D1}) and (\ref{D2})
we obtain

\begin{eqnarray}\label{Gamma_MOM}
&& \widetilde\gamma_{\mbox{\scriptsize MOM}}(\alpha) =
-\frac{\alpha}{\pi} + \frac{\alpha^2(1 + N_f)}{2\pi^2} +
O(\alpha^3);\\
&&\label{Beta_MOM} \widetilde\beta_{\mbox{\scriptsize
MOM}}(\alpha) = \frac{\alpha^2 N_f}{\pi}\Big(1+ \frac{\alpha}{\pi}
- \frac{\alpha^2}{2\pi^2}\Big(1 + 3 N_f
\left(1-\zeta(3)\right)\Big) + O(\alpha^3)\Big).\qquad
\end{eqnarray}

\noindent Comparing these equations we see that in the MOM scheme
only terms proportional to $(N_f)^1$ satisfy the NSVZ relation.
Note that the $\beta$-function in the MOM subtraction scheme
coincides with the Gell-Mann--Low function \cite{Gorishnii:1990kd}
and should not depend on the regularization. The same statement is
obvious for the anomalous dimension in the MOM scheme. The RG
functions (\ref{Gamma_MOM}) and (\ref{Beta_MOM}) are obtained
using the HD regularization. We have also verified these
expressions by the calculation of the anomalous dimension and the
$\beta$-function in the MOM scheme using the DRED regularization.
(In the three-loop approximation we have evaluated  only the
scheme-dependent terms proportional to $(N_f)^2$.) The results
coincide with Eqs. (\ref{Gamma_MOM}) and (\ref{Beta_MOM}). This
confirms the correctness of the calculations made with the HD
regularization.

A three-loop $\beta$-function and a two-loop anomalous dimension
for a general ${\cal N}=1$ SYM theory with matter in the
$\overline{\mbox{DR}}$-scheme have been calculated in Ref.
\cite{Jack:1996vg}.\footnote{In order to obtain the results of
Ref. \cite{Jack:1996vg} it is necessary to set $\alpha =
g^2/4\pi$, $\gamma(\alpha) = 2\gamma(g)$, $\beta(\alpha) = g
\beta(g)/2\pi$.} The result has the following form:

\begin{eqnarray}\label{Gamma_DRED}
&& \widetilde\gamma_{\overline{\mbox{\scriptsize{DR}}}}(\alpha) =
-\frac{\alpha}{\pi} + \frac{\alpha^2(1+N_f)}{2\pi^2} +
O(\alpha^3);\\
&&\label{Beta_DRED}
\widetilde\beta_{\overline{\mbox{\scriptsize{DR}}}}(\alpha) =
\frac{\alpha^2 N_f}{\pi}\Big(1 + \frac{\alpha}{\pi} -
\frac{\alpha^2(2+3N_f)}{4\pi^2} + O(\alpha^3)\Big).
\end{eqnarray}

\noindent Comparing these RG functions we see that the NSVZ
relation is valid for the terms proportional to $(N_f)^1$ and is
not satisfied for terms proportional to $(N_f)^2$. All terms
proportional to $(N_f)^0$ in different expressions for the
anomalous dimension coincide. Similarly, all terms proportional to
$(N_f)^1$ in different expressions for the $\beta$-function also
coincide. This confirms the general conclusions made in this
paper. Also we note that $\zeta(3)$ is present in the three-loop
$\beta$-function in the MOM scheme and is absent in the expression
found with the $\overline{\mbox{DR}}$ scheme exactly as in the
usual quantum electrodynamics in the MOM and
$\overline{\mbox{MS}}$ schemes, respectively (see e.g.
\cite{Gorishnii:1990kd}). The reason is that in both theories a
certain finite scalar integral proportional to $\zeta(3)$
\cite{Chetyrkin:1980pr} is essential, if the three-loop
$\beta$-function is calculated in the MOM scheme. Also it is
interesting to note that the anomalous dimension in the MOM scheme
coincides with the one in the $\overline{\mbox{DR}}$ scheme.

\section{Conclusion}
\hspace{\parindent}

For ${\cal N}=1$ SQED with $N_f$ flavors the exact NSVZ
$\beta$-function is obtained for the RG functions defined in terms
of the bare coupling constant if the theory is regularized by
higher derivatives. These RG functions by definition do not depend
on a choice of the renormalization scheme. However, the RG
functions defined in terms of the renormalized coupling constant
depend on a subtraction scheme. In this paper we have demonstrated
that the coefficients of the $\beta$-function proportional to
$(N_f)^1$ are scheme independent and satisfy the NSVZ relation in
all orders. This is explicitly verified by calculating the
two-loop anomalous dimension and the three-loop $\beta$-function
using different subtraction schemes.

\bigskip
\bigskip

\noindent {\Large\bf Acknowledgements.}

\bigskip

\noindent The  work of one of us (AK) is   supported  in part by
the  Grant NSh-2835.2014.2, Russian Foundation of Basic Research
grants N 11-01-00182 and N 11-02-00112. The work of KS was
supported by Russian Foundation for Basic Research grant No
11-01-00296.


\end{document}